\documentclass[aps,prx,twocolumn,amsmath,amssymb,superscriptaddress,floatfix]{revtex4-2}

\usepackage{mathtools}
\usepackage{multirow}
\usepackage{xcolor}
\usepackage{bm}
\usepackage{amsfonts,amssymb,amsmath}
\usepackage{graphicx,dcolumn,bm,xcolor,braket,slashed}
\usepackage{times} 
\usepackage{comment}
\usepackage{array}
\usepackage{textcomp}
\usepackage[normalem]{ulem}
\usepackage{dsfont}

\usepackage[title,toc,titletoc,page]{appendix}

\newcommand{\be}{\begin{eqnarray}}
\newcommand{\ee}{\end{eqnarray}}
\newcommand{\bbm}{\begin{bmatrix}}
\newcommand{\ebm}{\end{bmatrix}}
\newcommand{\bpm}{\begin{pmatrix}}
\newcommand{\epm}{\end{pmatrix}}

\usepackage{bibunits}
\defaultbibliographystyle{unsrt} 

\begin{document}
\begin{bibunit}

\title{Analogs of spontaneous emission and lasing in photonic time crystals}

\author{Kyungmin Lee}
\affiliation{Department of Physics, Korea Advanced Institute of Science and Technology, Daejeon 34141, Republic of Korea}

\author{Minwook Kyung}
\affiliation{Department of Physics, Korea Advanced Institute of Science and Technology, Daejeon 34141, Republic of Korea}

\author{Yung Kim}
\affiliation{Department of Physics, Korea Advanced Institute of Science and Technology, Daejeon 34141, Republic of Korea}

\author{Jagang Park}
\affiliation{Department of Electrical Engineering and Computer Sciences, University of California, Berkeley, California 94720, USA}

\author{Hansuek Lee}
\affiliation{Department of Physics, Korea Advanced Institute of Science and Technology, Daejeon 34141, Republic of Korea}

\author{Joonhee Choi}
\affiliation{Department of Electrical Engineering, Stanford University, Stanford, CA 94305, USA}

\author{C. T. Chan}
\affiliation{Department of Physics, the Hong Kong University of Science and Technology, Clear Water Bay, Kowloon, Hong Kong 999077, China}

\author{Jonghwa Shin}
\affiliation{Department of Material Sciences and Engineering, Korea Advanced Institute of Science and Technology, Daejeon 34141, Republic of Korea
}

\author{Kun Woo Kim}
\email{kunx@cau.ac.kr}
\affiliation{Department of Physics, Chung-Ang University, 06974 Seoul, Republic of Korea}

\author{Bumki Min}
\email{bmin@kaist.ac.kr}
\affiliation{Department of Physics, Korea Advanced Institute of Science and Technology, Daejeon 34141, Republic of Korea}

\begin{abstract}
We report the first direct mapping of the frequency‑resolved local density of states (LDOS) in a photonic time crystal (PTC) implemented as an array of time‑periodically modulated LC resonators at microwave frequencies. Broadband white noise probes the system and yields an LDOS lineshape near the momentum gap that can be decomposed into absorptive and dispersive Lorentzian components. The finite LDOS peak at the gap frequency, which grows with modulation strength, implies that the spontaneous emission rate of an emitter coupled to the PTC would be maximized at that frequency. The measured spectra are in good agreement with classical non‑Hermitian Floquet theory. As the modulation‑induced gain exceeds intrinsic losses, the system undergoes a transition to a narrow-band self-oscillation (lasing) regime. These results open a route to nonequilibrium photonics and bring time‑periodic LDOS engineering closer to practical realization.
\end{abstract}

\maketitle
\newpage

Controlling spontaneous emission rates by reshaping the electromagnetic vacuum has been a central theme in nanophotonics since Purcell’s pioneering cavity proposal~\cite{PhysRev.69.674.2}, and has been pursued in photonic crystals~\cite{1972JETP...35..269B,PhysRevLett.58.2059,john1987strong,fan1997high,PhysRevLett.95.013904,PhysRevLett.99.023902,PhysRevLett.99.193901,Noda2007} and, more recently, hyperbolic metamaterials~\cite{jacob2012broadband,noginov2010controlling}. In these platforms, within the weak‑coupling regime, the spontaneous emission rate of a dipole emitter is governed by the local photonic density of states (LDOS), which quantifies the vacuum field fluctuations available to a quantum emitter at a given frequency, position, and dipole orientation~\cite{lodahl2004controlling,vos2009orientation,novotny2012principles,milonni2013quantum}. Although static, passive nanostructures can shape the LDOS in both space and frequency, their modal spectrum is fixed by geometry and material parameters once fabricated.

Photonic time crystals (PTCs) break this constraint by modulating the optical parameters such as refractive index or permittivity in time rather than relying on spatial structuring~\cite{zurita2009reflection,salem2015temporal,PhysRevA.93.063813,doi:10.1063/1.4928659, PhysRevB.98.085142,martinez2018parametric, Chamanara2018, Park2021, Lee2021,asgari2024theory_,wang2024expanding,sustaeta2025quantum,dong2025extremely}. Note that our use of the term \textit{time crystal} differs from the conventional notion of a discrete time crystal, which refers to a many‑body quantum or classical phase that spontaneously breaks discrete time‑translation symmetry. A periodic modulation at frequency~$\Omega$ folds the dispersion into Floquet replicas that hybridize positive‑ and negative‑frequency branches and open momentum gaps~\cite{doi:10.1126/sciadv.abo6220,doi:10.1126/sciadv.adg7541,feinberg2025plasmonic}. Classical electrodynamics predicts that, near the edges of a momentum gap, the Petermann factor, which quantifies Floquet-mode non-orthogonality, enhances the contributions of individual modes to the momentum-resolved density of states (kDOS). Interestingly, in parameter regimes with net time-periodicity-induced gain, the kDOS can become negative, foreshadowing unconventional effects such as spontaneous emission excitation~\cite{park2025spontaneous}. These predictions are based on a classical non-Hermitian Maxwell–Floquet framework and can also be inferred from a quantum electrodynamical description~\cite{bae2025cavity}, although the corresponding kDOS and LDOS have not yet been directly tested experimentally.

In this Letter, we report what is, to our knowledge, the first direct measurement of the spectrally resolved LDOS and, by extension, the spontaneous‑emission spectrum of an emitter coupled to a PTC. Operating in the linear but lossy regime, we measure the radiated power generated by broadband white noise in the driven circuits that constitute the PTC. The resulting spectrum exhibits a pronounced cusp at the momentum gap frequency and can be decomposed into absorptive and dispersive Lorentzian components. All features agree well with classical non‑Hermitian Floquet theory, which predicts an enhancement of the LDOS, and therefore of the spontaneous‑emission decay rate, at the gap frequency~\cite{park2025spontaneous}. These measurements provide the first direct evidence that purely temporal modulation can sculpt the electromagnetic environment of a light emitter. Furthermore, above a threshold modulation power we observe a transition to a narrow-band oscillation consistent with modulation-induced parametric self-oscillation (hereafter a \textit{PTC laser}).

\begin{figure}[htb!]
  \centering
    \includegraphics[width=0.45\textwidth]{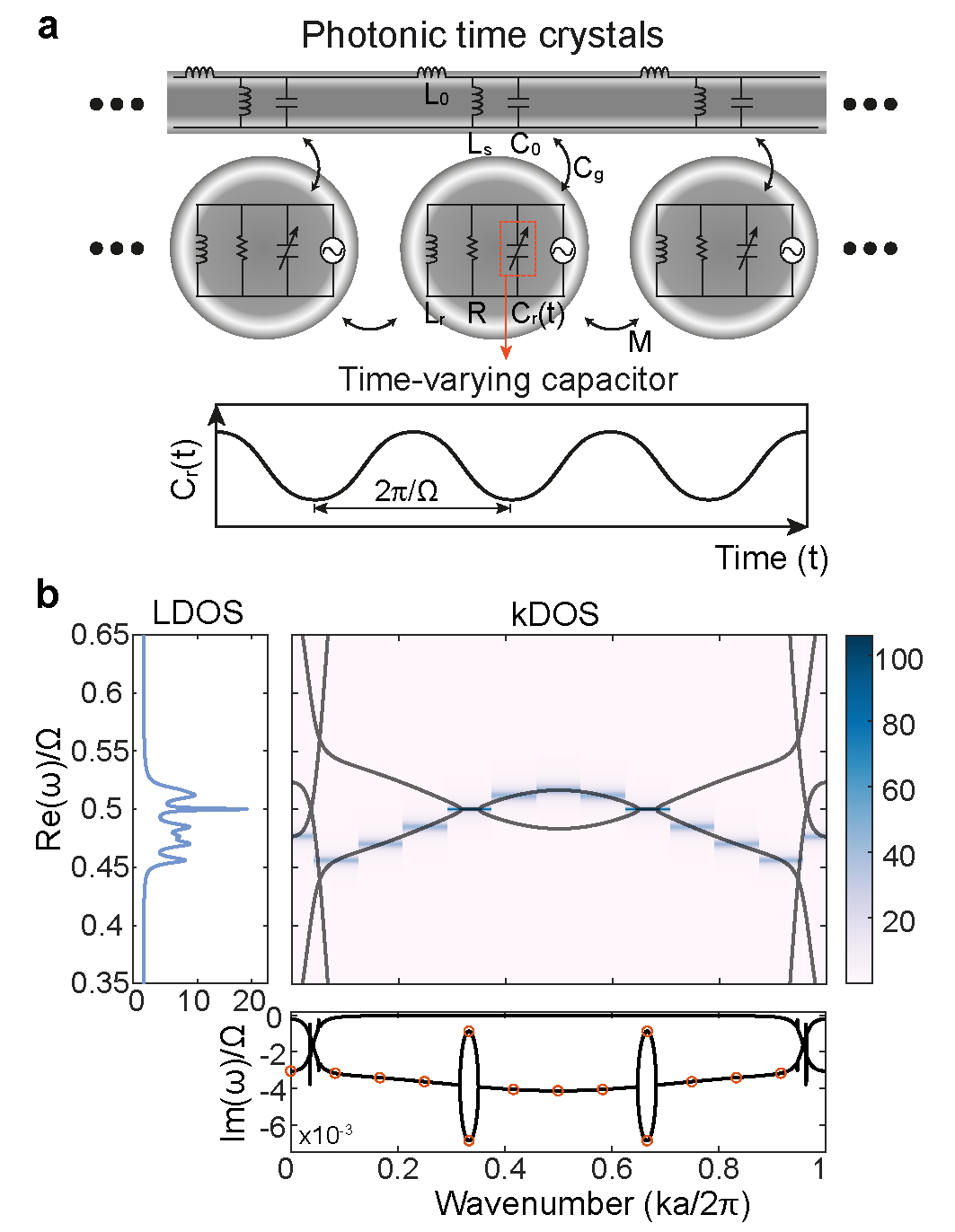}
  \caption{\label{fig:Fig1} (a) Schematic representation of the equivalent circuit for the photonic time crystal. The system consists of an array of LC resonators side‑coupled to a transmission line. In each unit cell, the resonator capacitors are sinusoidally modulated at angular frequency $\Omega$, thereby producing a time‑periodic modulation of the effective permittivity of the structure. (b) Map of the momentum-resolved density of states (kDOS) together with band structures (main panel), LDOS (left panel), and imaginary parts of the quasi-eigenfrequencies (bottom panel) for $\Omega \approx 2\omega_{k}$ at $\delta = 0.023$.}
\end{figure}

We first analyze the band structure of the PTC implemented as a one‑dimensional array of LC resonators side‑coupled to a bus waveguide (Fig.~\ref{fig:Fig1}a). The capacitance of each resonator is modulated periodically in time as $C_r(t)=C_c[1+\delta(t)]^{-1}$, where $\delta(t)$ is a dimensionless function with period $T = 2\pi/\Omega$. In momentum space, the complex amplitudes of the waveguide and resonator modes, $a_k$ and $b_k$, evolve under a time‑dependent Hamiltonian $ \mathcal{H}(t)=\mathcal{H}^{(0)}+\mathcal{H}^{(1)}(t)$. The static part $\mathcal{H}^{(0)}$ contains the bare dispersions of the waveguide and resonators and their mutual coupling, whereas the time-dependent part $\mathcal{H}^{(1)}(t)$ originates from the periodic capacitance drive. Defining the mode vector as $\mathbf{x}_k=[a_k,b_k,a_{-k}^\dagger,b_{-k}^\dagger]^{T}$, the equations of motion reduce to the compact form $i\dot{\mathbf{x}}_k=\mathbf{H}_k(t)\mathbf{x}_k$, with $\mathbf{H}_k(t)=\mathbf{H}_k^{(0)}+\mathbf{H}_k^{(1)}(t)+\mathbf{D}_k$. To model the lossy regime, each resonator is equipped with a resistor, which leads to a loss-induced term $\mathbf{D}_k$ in the dynamical matrix.  Full expressions are provided in Supplementary Material B.

For simplicity, we assume a single-harmonic modulation $\delta(t) = \delta \cos(\Omega t)$, where $\Omega$ is the modulation frequency. Under this assumption, the Floquet Hamiltonian $\mathbf{H}_k^{F}$ is block-tridiagonal. The corresponding eigenvalue equation $\mathbf{H}_k^{F}\mathbf{F}_{k,\alpha}
= \omega_\alpha\mathbf{F}_{k,\alpha}$, where $\mathbf{F}_{k,\alpha}$ denotes the Floquet eigenvector, defines the quasi-frequency band structure of the PTC. At the near-resonant condition $\Omega\simeq2\omega_k$, with $\omega_k$ representing the Bloch‑mode frequency of the resonator array, asymmetric coupling occurs between the original positive-frequency band and the Floquet sideband of its corresponding negative-frequency band, opening a momentum gap around $\Omega/2$ (Fig.~\ref{fig:Fig1}b). The plot combines continuum-limit Bloch–Floquet dispersions with discrete $k$-resolved resonances whose spacing reflects the finite number of unit cells ($N=12$). At the edges of this gap, we observe exceptional points (EPs), where Floquet eigenmodes and their complex quasi-eigenfrequencies coalesce. Within the momentum gap, the two modes exhibit unequal decay rates, as reflected by the splitting in the imaginary parts of the quasi-eigenfrequencies (see the bottom panel of Fig. \ref{fig:Fig1}b).
\begin{figure}
  \centering
    \includegraphics[width=0.45\textwidth]{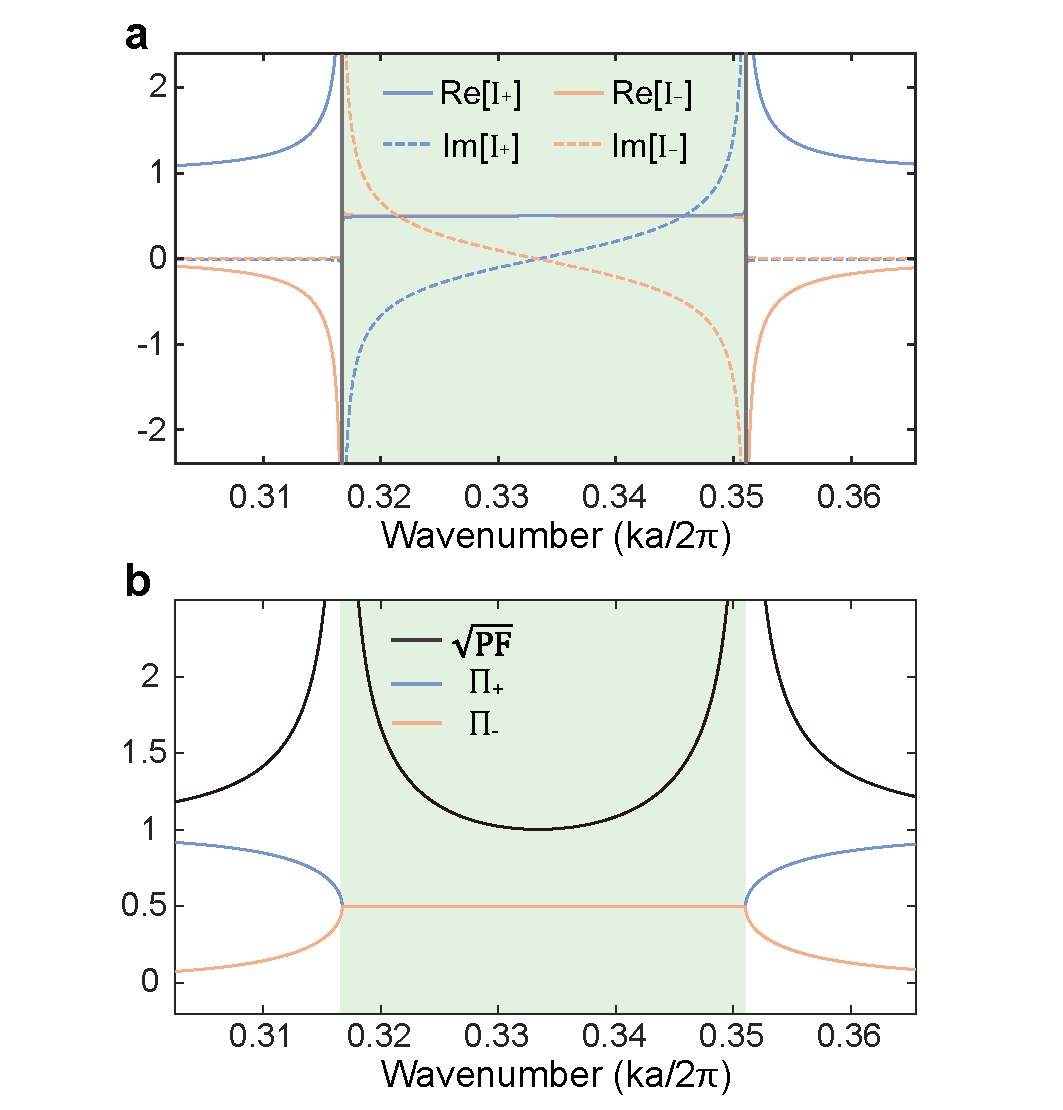}
  \caption{\label{fig:Fig2}
  (a) Real (solid lines) and imaginary (dashed lines) parts of the normalized field intensity $I_\pm(k)$. Inside the momentum gap (green shaded region), a finite $\Im[I_\pm(k)]$ generates the antisymmetric Lorentzian component in the kDOS. Outside the gap, $\Im[I_\pm(k)] \approx 0$, resulting in a purely symmetric Lorentzian profile. (b) Square root of the Petermann factor, $\sqrt{\mathrm{PF}}=\sqrt{\mathrm{PF}_+}=\sqrt{\mathrm{PF_-}}$ (black), and source-mode overlap magnitude $\Pi_\alpha$ for the two modes ($+$ in blue and $-$ in red)}
\end{figure}

In the Floquet picture, the response of the system is governed by the Floquet Green's function $G_F(k,\omega) = [\omega \mathbf{I}_F-\mathbf{H}_k^F]^{-1}$, where $\mathbf{H}_k^F$ contains all harmonic blocks ($q=0,\, \pm1,\,\pm2,\,\dots$) and loss. A monochromatic drive at frequency $\omega$ couples to the resonator mode $b_k$ through $\mathbf{s}_0 = [0,-1,0,1]^T$ in the $q=0$ block, thereby exciting all sideband components $b_k^{(q)}$ at $\omega+q\Omega$. Embedding $\mathbf{s}_0$ in the full Floquet space,
$\mathbf{s}=[\dots;\mathbf{0};\mathbf{s}_0;\mathbf{0};\dots]$, we project the Floquet Green's function onto the same ($b_k,q=0$) channel:
\begin{equation}
    g_F(k,\omega)=\mathbf{s}^T G_F(k,\omega)\mathbf{s},
\end{equation}
which is the Green's function for the resonator mode $b_k$ at frequency $\omega$. Here, $g_F$ describes the flux response; dividing by the $k$-dependent effective inductance $L'_{r,k}$ (see Supplementary Material B) converts it into the complex current-to-voltage response $Y_F(k,\omega) = g_F(k,\omega)/L'_{r,k}$, i.e., the admittance at $(k,\omega)$. The poles $\omega_\alpha=\Omega_\alpha -i\Gamma_\alpha$ reproduce the quasi-eigenfrequencies, while $\Im [Y_F]$ gives the spectral weight that can absorb power from the drive. We therefore define the momentum-resolved density of states (kDOS)
\begin{equation}
    \rho_k(\omega) \equiv \frac{2}{\pi}\,
     \Im\bigl[Y_{F}(k,\omega)\bigr]=\frac{2}{\pi}\frac{1}{L'_{r,k}}\,
     \Im\bigl[g_{F}(k,\omega)\bigr]
\end{equation}
and obtain the LDOS by summation $\rho(\omega)=N^{-1}\sum_k\rho_k(\omega)$. Neglecting edge effects under the bulk approximation, we treat the LDOS as translationally invariant and omit the site index. The LDOS shows a sharp peak at the momentum gap frequency (Fig.~\ref{fig:Fig1}b), confirming our recent prediction that spontaneous emission decay rates are enhanced at this spectral point~\cite{park2025spontaneous}. If we consider only the two modes $\omega_\pm=\Omega_\pm -i\Gamma_\pm$ forming the momentum gap, the kDOS can be approximately decomposed as
\begin{equation}
\begin{aligned}
    \rho_k(\omega) &\approx \sum_{\alpha=\pm} \rho_k^\alpha(\omega)\\
    &\approx\sum_{\alpha=\pm}\frac{1}{\pi}\frac{\Gamma_\alpha}{(\omega-\Omega_\alpha )^2+\Gamma_\alpha^2}\frac{\Re[I_\alpha (k)]}{L'_{r,k}}\\
    &\qquad\,+\frac{1}{\pi}\frac{\Omega_\alpha-\omega}{(\omega-\Omega_\alpha)^2+\Gamma_\alpha^2}\frac{\Im[I_\alpha (k)]}{L'_{r,k}},
\end{aligned} \label{eqn:mode decomposition}
\end{equation}
where the normalized field intensity $I_\alpha (k)$ is expressed as 
\begin{equation}
    I_\alpha (k) = \frac{2\mathbf{s}^T\mathbf{F}_{k,\alpha}^R\mathbf{F}_{k,\alpha}^L\mathbf{s}}{\mathbf{F}_{k,\alpha}^L\mathbf{F}_{k,\alpha}^R}
\end{equation}
with $\mathbf{F}_{k,\alpha}^R$ and $\mathbf{F}_{k,\alpha}^L$ denoting the right and left eigenvectors of the Floquet Hamiltonian $\mathbf{H}_k^F$, respectively. Since the denominator involves the overlap between left and right eigenvectors, the magnitude $|I_\alpha (k)|$ grows with the square root of the Petermann factor, capturing the effect of non-orthogonal eigenmodes. The real and imaginary parts of $I_\alpha (k)$ determine the weights of the symmetric and antisymmetric Lorentzian components that constitute the kDOS. For momenta $k$ within the band region (i.e., outside the momentum gap), $\Im[I_\alpha (k)]$ is negligible, yielding a symmetric Lorentzian profile. In contrast, inside the momentum gap, $\Im[I_\alpha (k)]$ becomes significant, introducing an antisymmetric contribution to the kDOS (Fig.~\ref{fig:Fig2}a). The weight of the antisymmetric component increases as the modes approach the gap edges and vanishes at the gap center.

To further clarify the contribution of non-orthogonality to the modal weights, Fig.~\ref{fig:Fig2}b plots the square root of the Petermann factor $\sqrt{\mathrm{PF}_\alpha}$ and the source–mode overlap $\Pi_\alpha$. As shown, the overlap magnitude $\Pi_\alpha \equiv |\mathbf{s}^T\mathbf{F}_{k,\alpha}^R\mathbf{F}_{k,\alpha}^L\mathbf{s}|$ remains nearly constant throughout the momentum gap (shaded region), whereas $\sqrt{\mathrm{PF}_\alpha}$ sharply increases toward the gap edges. Because the normalized intensity satisfies
\begin{equation}
    |I_\alpha| = 2\sqrt{\mathrm{PF}_\alpha}\,\Pi_\alpha,
\end{equation}
the variation of $|I_\alpha(k)|$ inside the gap (Fig.~\ref{fig:Fig2}a) directly reflects the increase of $\sqrt{\mathrm{PF}_\alpha}$. Thus, the enhancement of the antisymmetric contribution to the kDOS originates from the growing non-orthogonality of the two in-gap modes near the exceptional points at the edges of the momentum gap.

To implement the unit cell of the PTC, a metallic split-ring resonator (SRR) was patterned onto a printed circuit
board. Within this unit cell, we incorporated a varactor—an element whose capacitance changes when the voltage across it is varied. By applying a DC-biased AC voltage to the varactor, the resonant characteristic of the unit cell can be time-periodically modulated. To build a PTC, we inserted an array of 12 unit cells into the waveguide. Further details of the fabricated LC resonator and the experimental setup are provided in Supplementary Material D. To verify the predicted LDOS enhancement at the momentum-gap frequency, we operate the PTC with no coherent input; the only drive is broadband white noise arising from Johnson–Nyquist thermal fluctuations together with residual technical noise. These fluctuations act as spatially distributed, broadband noise sources that probe the linear response of the PTC. The measured noise-driven radiated spectrum directly follows the LDOS lineshape predicted by non-Hermitian Floquet theory, providing an experimental proxy for the frequency dependence of the spontaneous-emission rate in the weak-coupling limit.

\begin{figure*}[htb!]
  \centering
    \includegraphics[width=0.9\textwidth]{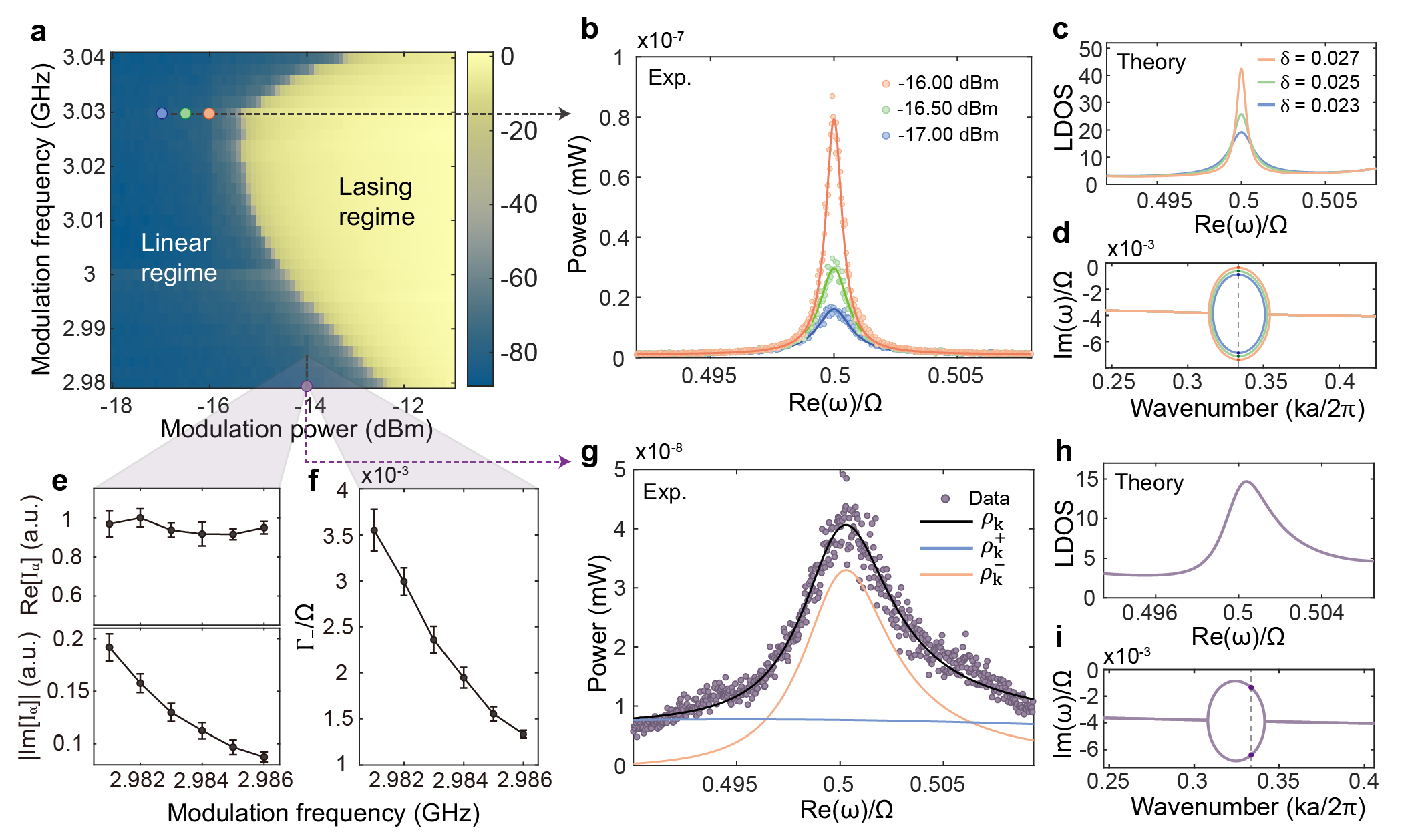}
  \caption{\label{fig:Fig3} (a) Experimentally measured radiated-power map at the momentum gap frequency as functions of modulation frequency and power. Blue (linear) and yellow (oscillation/lasing) regions are separated by the oscillation threshold; colored markers indicate operating points used in the panels b and g. (b) Sub-threshold spectra measured for the gap-center modes. Increasing the modulation power from -17 dBm to -16 dBm increases the peak amplitude and narrows the symmetric Lorentzian profile. (c) The calculated LDOS for modulation depths in the range $0.023 \le \delta \le 0.027$ reproduces the observed peak growth and linewidth narrowing. (d) The imaginary part of the eigenfrequencies as a function of wavenumber $k$. The two gap modes dominate the LDOS at the momentum gap frequency. (e) Real and imaginary parts of $I_\alpha $ for the gap modes as a function of modulation frequency. The imaginary parts of $I_\alpha $ decrease as the mode approaches the center of the momentum gap, reducing the antisymmetric weight in the LDOS. (f) Extracted linewidth $\Gamma_-=\Im[\omega_-]$ of the lower-loss gap mode. The linewidth decreases toward the gap center, accompanied by a narrowing of the LDOS peak.  (g) When the modulation frequency is detuned so the modes move off-center in the momentum gap, the measured spectrum becomes skewed. Black curve: total fit obtained from the two-mode model of Eq. \eqref{eqn:mode decomposition}. Blue and orange curves: individual contributions of the  $\alpha=+$ and $-$ gap modes, respectively. (h) Theoretical LDOS for this detuned case reproduces the asymmetric profile. (i) Corresponding imaginary part of the eigenfrequencies versus wavenumber $k$.}
\end{figure*}

Figure~\ref{fig:Fig3}a shows an experimentally measured map of the oscillation power at the momentum gap frequency as a function of the modulation power and frequency. Once the modulation crosses threshold, the array exhibits self-oscillation; in the above-threshold regime the varactor nonlinearity becomes essential and leads to saturation, beyond linear Floquet theory. We begin by analyzing the linear regime, where the Floquet description remains valid. To ensure operation within this linear, lossy response regime, we systematically reduced the modulation power below the oscillation threshold and recorded the radiated power spectral density, which corresponds to the spectrally resolved LDOS (Fig.~\ref{fig:Fig3}b, g). 
\begin{figure}
  \centering
    \includegraphics[width=0.45\textwidth]{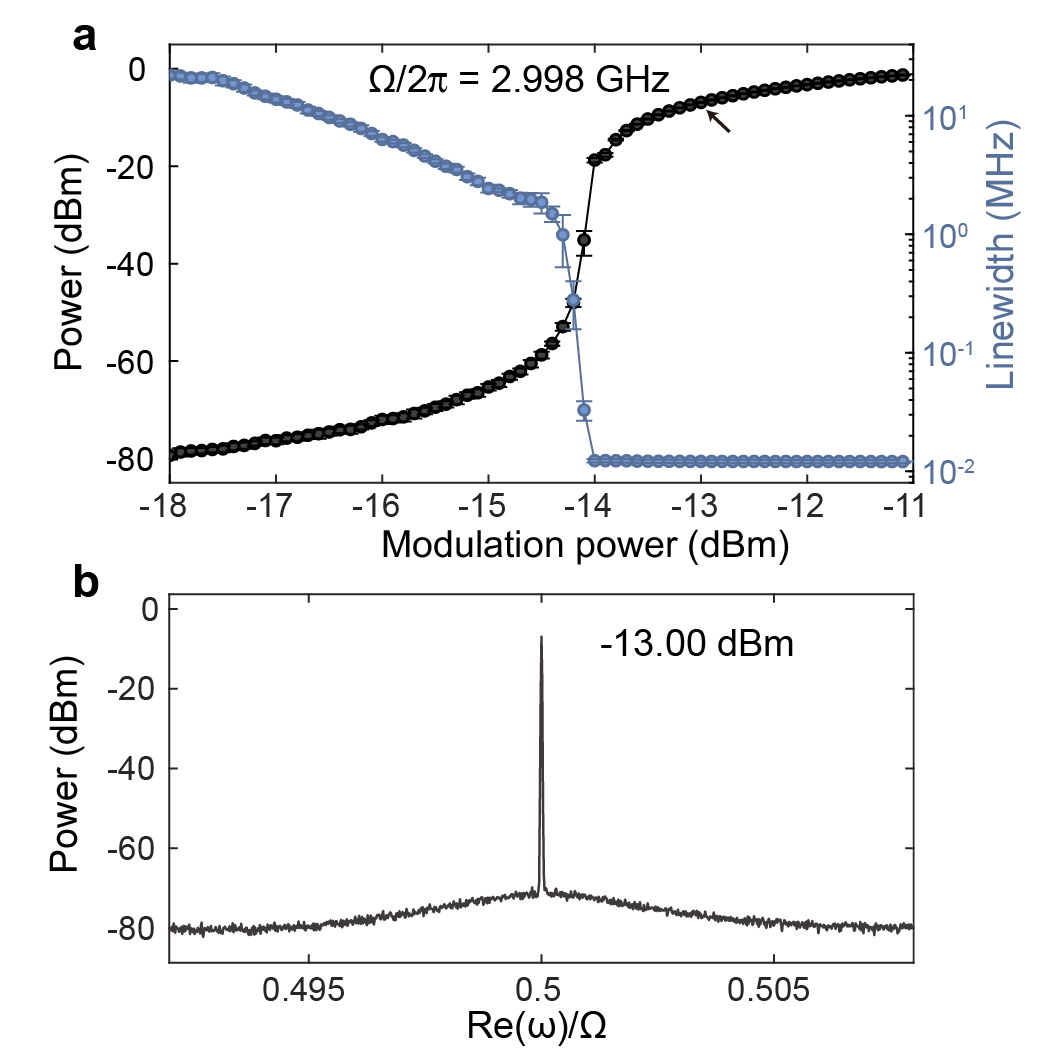}
  \caption{\label{fig:Fig4} (a) Measured radiated power at the momentum gap frequency and the corresponding linewidth as functions of modulation power at a fixed modulation frequency $\Omega/2\pi=2.998 \,\mathrm{GHz}$. As the modulation power exceeds the threshold, the system undergoes a sharp transition into an oscillating regime where the radiated power saturates due to nonlinearities. (b) Spectral profile at $-13 \,\mathrm{dBm}$, deep in the lasing regime, showing a sharp, narrow-band oscillation. }
\end{figure}

In the experiment, the position of the modes within the momentum gap can be tuned by adjusting the modulation frequency $\Omega$. We begin with the case where $\Omega$ is set so that the modes lie at the center of the gap.
With this frequency fixed, we measure the spectrum while sweeping the modulation power by adjusting the peak-to-peak AC voltage applied to the varactor diodes, each biased at a fixed reverse DC voltage. The measured spectra in Fig.~\ref{fig:Fig3}b are well fitted by a symmetric Lorentzian function, with increasing modulation power leading to a higher peak amplitude and a narrower linewidth. This behavior is closely reproduced by the theoretical curves in Figs.~\ref{fig:Fig3}c,d. When the modes are displaced from the center of the momentum gap, the measured LDOS becomes noticeably asymmetric with respect to the gap frequency. This skewness is fully consistent with our prediction that the spectrum consists of symmetric and antisymmetric Lorentzian components (Figs.~\ref{fig:Fig3}g–i).

To quantitatively capture the evolving lineshape within the momentum gap, we analyze the behavior of the key parameters in Eq.~\eqref{eqn:mode decomposition}: the real and imaginary parts of $I_\alpha$, which control the
weights of the symmetric and antisymmetric components, and the mode linewidth $\Gamma_\alpha$. These quantities are extracted by fitting the measured spectra to the two-mode expression. Figure~\ref{fig:Fig3}e shows that $|\Im[I_\alpha]|$ decreases with increasing modulation frequency, as the mode approaches the center of the gap, consistent with the transition from a skewed to a symmetric spectral profile. In contrast, $\Re[I_\alpha]$ remains nearly constant throughout. Among the two gap modes, the one with the larger linewidth, $\Gamma_+$, is strongly damped and buried beneath the noise floor, making reliable extraction of its parameters difficult. We therefore focus on the less-damped mode, whose linewidth is $\Gamma_-$, as its spectral contribution remains clearly visible across the modulation range. As the modulation frequency increases, the gap mode moves closer to the center of the momentum gap, where the imaginary part of its eigenfrequency—and thus the linewidth—tends to decrease
(Fig.~\ref{fig:Fig3}f). This trend is consistent with the characteristic band structure of a PTC, where the gap modes exhibit minimal damping near the gap center and stronger attenuation toward the edges.

Once the modulation-induced parametric gain overcomes the losses of the finite-sized PTC, the in-gap Floquet mode becomes unstable and the system enters a self-oscillating (parametric-oscillator) regime (Fig.~\ref{fig:Fig4}). Above threshold, a coherent narrow-band oscillation appears; we call this regime a “PTC laser” to emphasize its laser-like coherence despite its parametric origin, and it saturates due to varactor nonlinearities. Unlike conventional population-inversion lasers, the amplification here is supplied by the external time-periodic modulation and is therefore best viewed as modulation-induced parametric gain rather than material gain. We therefore interpret the oscillation as modulation-induced parametric self-oscillation, closely analogous to a degenerate optical parametric oscillator~\cite{eckardt1991optical}. The observed threshold is consistent with the onset of a parametric instability of the in-gap mode, i.e., when the imaginary part of its quasi-eigenfrequency approaches zero ($\Gamma_- \to 0$) within the linear Floquet model. Within a near-resonant, two-mode Floquet approximation, the instability occurs when the modulation depth reaches the critical value
\begin{equation}
    \delta_c = \frac{4}{\omega_k}\sqrt{\Delta^2+\gamma^2},
\end{equation}
where $\Delta \equiv \Omega/2 - \omega_k$ is the detuning from exact parametric resonance and $\gamma$ denotes the damping rate of the resonator Bloch mode (see Supplementary Material B for details). Beyond threshold, the oscillation amplitude grows to a point where the linear Floquet description breaks down: the nonlinear response of
the varactors becomes essential, rendering the earlier LDOS analysis inapplicable. In this strongly driven regime, the PTC behaves effectively as a one-dimensional array of nonlinear, parametrically driven LC oscillators. From the viewpoint of coupled nonlinear parametric oscillators, such systems are natural candidates for activated discrete-time-crystalline behavior, exhibiting synchronized subharmonic oscillations that spontaneously break the discrete time-translation symmetry of the drive~\cite{yao2020classical}. While distinct from the photonic time crystal considered here, discrete time-crystalline phases have attracted wide interest across classical and quantum platforms~\cite{PhysRevLett.123.124301,pizzi2021aclassical,PhysRevE.100.060105,taheri2022all,giergiel2023discrete,PhysRevLett.127.140603,yi2024theory,choi2017observation,zhang2017observation,ho2017critical,rovny2018p,liu2023photonic,kyprianidis2021observation,mi2022time}.

We have shown that a PTC, implemented here with an array of varactor‑loaded microwave resonators, can tailor the electromagnetic vacuum in a manner analogous to spatial structuring, but using time as the control knob. By measuring the noise-seeded radiated power spectral density emitted from the unit-cell circuits, we directly mapped the spectrally resolved LDOS and observed it reshape as we varied the drive frequency and modulation depth. The measured lineshape, its skewness, and the steep growth of the peak all follow the predictions of non‑Hermitian Floquet theory, including the Petermann factor enhancement associated with exceptional points at the momentum‑gap edges. These experiments therefore place the LDOS—arguably the most fundamental quantity in light–matter interaction—squarely under active temporal control. Looking ahead, this approach opens a route to dynamic Purcell engineering, tabletop tests of dynamical Casimir analogs, and, in suitably coupled platforms, on‑demand switching between emission, absorption, and even spontaneous excitation regimes for quantum emitters interfaced with photonic time crystals.

\noindent \textit{Acknowledgement.---} This work is supported by the National Research Foundation of Korea (NRF) through the government of Korea (NRF-2022R1A2C301335313) and the Samsung Science and Technology Foundation (SSTF-BA2402-02). 
K.W.K. acknowledges financial support from the Basic Science Research Program through the National Research Foundation of Korea (NRF) funded by the Ministry of Education (no. RS-2025-00521598) and the Korean Government (MSIT) (no. 2020R1A5A1016518).
J.C. acknowledges support from the Terman Faculty Fellowship at Stanford University. 
Work in Hong Kong was supported by the Research Grants Council (Grant No. AoE/P-502/20).


\putbib[refs]
\end{bibunit}

\onecolumngrid
 \clearpage
 
\begin{bibunit}

\begin{widetext}
\setcounter{equation}{0} 
\setcounter{figure}{0} 
\renewcommand{\theequation}{S\arabic{equation}} 
\renewcommand{\thefigure}{S\arabic{figure}} 

{\centering
\large \textbf{Supplemental Material: Analogs of spontaneous emission and lasing in photonic time crystals} \\[2ex]  

\normalsize Kyungmin Lee$^{1}$, Minwook Kyung$^{1}$, Yung Kim$^{1}$, Jagang Park$^{2}$, Hansuek Lee$^{1}$, Joonhee Choi$^{3}$, C. T. Chan$^{4}$, Jonghwa Shin$^{5}$, Kun Woo Kim$^{6,\dagger}$, Bumki Min$^{1,\ddagger}$ \\[1ex] 
\fontsize{9pt}{10pt}\selectfont $^{1}$\textit{Department of Physics, Korea Advanced Institute of Science and Technology, Daejeon 34141, Republic of Korea} \\
$^{2}$\textit{Department of Electrical Engineering and Computer Sciences, University of California, Berkeley, California 94720, USA} \\
$^{3}$\textit{Department of Electrical Engineering, Stanford University, Stanford, CA 94305, USA} \\
$^{4}$\textit{Department of Physics, the Hong Kong University of Science and Technology, Clear Water Bay, Kowloon, Hong Kong 999077, China} \\
$^{5}$\textit{Department of Material Sciences and Engineering, Korea Advanced Institute of Science and Technology, Daejeon 34141, Republic of Korea} \\
$^{6}$\textit{Department of Physics, Chung-Ang University, 06974 Seoul, Republic of Korea} \\
\textit{$^{\dagger}$kunx@cau.ac.kr} \\
\textit{$^{\ddagger}$bmin@kaist.ac.kr}\\
}

\renewcommand{\thesection}{\Alph{section}}
\counterwithin{figure}{section}

\section{Circuit model derivation of photonic time crystal Hamiltonian}

We model the photonic time crystal as a one-dimensional array of LC resonators side-coupled to a bus waveguide (Fig. 1a of the main text), as in our previous work\cite{park2022revealing}. The Lagrangian is 
\begin{equation}
\mathcal{L}=\sum_n\left(\frac{L_0}{2}(\dot{Q}_n^{L_0})^2+\frac{L_s}{2}(\dot{Q}_n^{L_s})^2+\frac{L_r}{2}(\dot{Q}_n^{L_r})^2+M\dot{Q}_n^{L_r}\dot{Q}_{n+1}^{L_r}-\frac{(Q_n^{C_0})^2}{2C_0}-\frac{(Q_n^{C_r})^2}{2C_r(t)}-\frac{Q_n^{C_r}Q_n^{C_0}}{C_g}\right)\label{eqn:Lag_position} 
\end{equation}
Here, \(Q_{n}^{X}\) denotes the charge on element \(X\) in unit cell \(n\), and \(\dot Q_{n}^{X}\) is the corresponding current. The lumped parameters $L_0$, $L_s$ and $C_0$ specify, respectively, the series inductance, shunt inductance, and ground capacitance that comprise the discretized waveguide. Each resonator contains an inductance \(L_{r}\) and a time-dependent capacitance \(C_{r}(t)\). The term $M$ denotes the mutual inductance between neighboring resonators, with only nearest-neighbor coupling assumed. The parameter $C_g$ represents the coupling capacitance between the resonator and the waveguide. Moving to momentum space, we label the Bloch modes by an integer $m=0,1,\dots,N-1$ and define the discrete wavenumbers $k_m \equiv 2\pi m/N$. The discrete Fourier transform is then
\begin{equation}
    x_{k_m}= \frac{1}{\sqrt{N}}\sum_{n=0}^{N-1}
         x_n\,e^{-ik_m n} ,
\qquad m=0,1,\dots,N-1 .
\end{equation}
For brevity, we henceforth write $k\equiv k_m$ and $\sum_k \equiv \sum_{m=0}^{N-1}$. For an open chain, the two missing end bonds act as boundary terms that mix different $k$-modes, but the corresponding matrix elements are suppressed by a factor of $\mathcal{O}(1/N)$. Accordingly, we work within the usual bulk approximation and disregard these boundary-induced mode couplings. Under this bulk approximation, the Lagrangian separates into independent $k$-sectors:
\begin{equation}
\begin{aligned}
    \mathcal{L}  =\sum_k\bigg[&\frac{L_s}{2}\bigg(\frac{L_s}{L_0}|s_k|^2+1\bigg)\dot{Q}_{-k}^{L_s}\dot{Q}_k^{L_s} +\bigg(\frac{L_r}{2}+M\cos{k}\bigg)\dot{Q}_{-k}^{L_r}\dot{Q}_k^{L_r}-\bigg(\frac{L_s}{L_0}|s_k|^2+1\bigg)^2\frac{Q_{-k}^{L_s}Q_k^{L_s}}{2C_0}\\
    & -\frac{Q_{-k}^{L_r}Q_{k}^{L_r}}{2C_r(t)}+\bigg(\frac{L_s}{L_0}|s_k|^2+1\bigg)\frac{Q_{-k}^{L_s}Q_k^{L_r}}{C_g}\bigg],
\end{aligned}
\end{equation}
where $s_k\equiv1-e^{ik}$. The conjugate node fluxes (canonical momenta) are
\begin{equation}
\begin{aligned}
    \Phi_k^{L_s} &= \frac{\partial\mathcal{L}}{\partial \dot{Q}_k^{L_s}}=L_s\big(\frac{L_s}{L_0}|s_k|^2+1\big)\dot{Q}_{-k}^{L_s},\\
     \Phi_k^{L_r} &= \frac{\partial\mathcal{L}}{\partial \dot{Q}_k^{L_r}} = L'_{r,k}\dot{Q}_{-k}^{L_r},
\end{aligned}
\end{equation}
where the $k$-dependent effective inductance is $L'_{r,k} = L_r+2M\cos{k}$. Applying a Legendre transformation to the Lagrangian gives the Hamiltonian in $k$-space:
\begin{equation}
\begin{aligned}
    \mathcal{H}&=\sum_k\sum_{\alpha=L_s,L_r}\frac{\partial \mathcal{L}}{\partial\dot{Q}_k^\alpha}\dot{Q}_k^\alpha-\mathcal{L}\\
    &=\sum_k\left(\frac{\Phi_{-k}^{L_s}\Phi_k^{L_s}}{2L_s^2C_0\Omega_k^2}+\frac{\Phi_{-k}^{L_r}\Phi_k^{L_r}}{2L_{r,k}'}+\frac{1}{2}L_s^2C_0\Omega_k^4Q_{-k}^{L_s}Q_k^{L_s}+\frac{Q_{-k}^{L_r}Q_k^{L_r}}{2C_r(t)}-L_sC_0\Omega_k^2\frac{Q_{-k}^{L_s}Q_k^{L_r}}{C_g}\right).
\end{aligned}
\end{equation}
where the waveguide dispersion is $\Omega_k = \sqrt{\frac{1}{L_sC_0}(\frac{L_s}{L_0}|s_k|^2+1)}$. We assume the resonator capacitance is modulated periodically in time, $C_r(t)=\frac{C_c}{1+\delta(t)}$, where $\delta(t)$ is a dimensionless, $T$-periodic function satisfying $\delta(t+T)=\delta(t)$. With this periodic modulation, the Hamiltonian separates into a static part and a time-dependent part:
\begin{equation}
    \mathcal{H}(t)=\mathcal{H}^{(0)}+ \mathcal{H}^{(1)}(t)
\end{equation}
where the static component is
\begin{equation}
    \mathcal{H}^{(0)}=\sum_k\bigg(\frac{\Phi_{-k}^{L_s}\Phi_k^{L_s}}{2L_s^2C_0\Omega_k^2}+\frac{\Phi_{-k}^{L_r}\Phi_k^{L_r}}{2L_{r,k}'}+\frac{1}{2}L_s^2C_0\Omega_k^4Q_{-k}^{L_s}Q_k^{L_s}+\frac{Q_{-k}^{L_r}Q_k^{L_r}}{2C_c}-L_sC_0\Omega_k^2\frac{Q_{-k}^{L_s}Q_k^{L_r}}{C_g}\bigg),
\end{equation}
and the time-dependent component is 
\begin{equation}
    \mathcal{H}^{(1)}(t)=\sum_k\frac{\delta(t)}{2C_c}Q_{-k}^{L_r}Q_k^{L_r}.
\end{equation}
We introduce the normal mode amplitudes for the waveguide and LC resonators
\begin{equation}
     a_k=\sqrt{\frac{L_s^2C_0\Omega_k^3}{2}}Q_k^{L_s}+i\sqrt{\frac{1}{2L_s^2C_0\Omega_k^3}}\Phi_k^{L_s},\; \, b_k = \sqrt{\frac{1}{2C_c\omega_k}}Q_k^{L_r}+i\sqrt{\frac{C_c\omega_k}{2}}\Phi_k^{L_r},
 \end{equation}
where $\omega_k = \sqrt{1/L'_{r,k}C_c}$ denotes the resonator array dispersion in the absence of modulation. Expressed in terms of these normal mode amplitudes, the static and time-dependent parts of the Hamiltonian become 
\begin{equation}
  \mathcal{H}^{(0)}=\sum_k\big[\Omega_ka_k^\dagger a_k+\omega_k b_k^\dagger b_k +g_k(a_{-k}+a_k^\dagger)(b_k+b_{-k}^\dagger)\big]
\end{equation}
and
\begin{equation}
  \mathcal{H}^{(1)}(t)=\sum_k\frac{\omega_k\delta(t)}{4}(b_{-k}+b_k^\dagger)(b_k+b_{-k}^\dagger)
\end{equation}
where the coupling strength between the waveguide and the resonators is given by $g_k=\sqrt{C_0C_c\Omega_k\omega_k}/(2C_g)$ .

\section{Floquet Hamiltonian of photonic time crystals}
For the mode vector $\mathbf{x}_k=[a_k,\,b_k,\,a_{-k}^\dagger,\,b_{-k}^\dagger]^T$, the static and time-dependent Hamiltonians, $\mathcal{H}^{(0)}$ and $ \mathcal{H}^{(1)}(t)$ can be written compactly: 
\begin{equation}
    \mathcal{H}^{(0)} = \sum_k\mathbf{x}_k^\dagger\hat{H}_k^{(0)}\mathbf{x}_k , \;\, \mathcal{H}^{(1)}(t)= \sum_k\mathbf{x}_k^\dagger\hat{H}^{(1)}_k(t)\mathbf{x}_k
\end{equation}
with the matrix representations
 \begin{equation}
    \hat{H}_k^{(0)} = \frac{1}{2}\begin{bmatrix}
        \Omega_k & -g_k & 0 & -g_k\\
        - g_k & \omega_k &  -g_k & 0\\
         0 &  -g_k & \Omega_k&-g_k\\
         -g_k & 0 & -g_k &\omega_k\end{bmatrix} , \,\; \hat{H}^{(1)}_k(t) =\frac{\omega_k\delta(t)}{4}\begin{bmatrix}
             0&0&0&0\\
             0&1&0&1\\
             0&0&0&0\\
             0&1&0&1
         \end{bmatrix}.
\end{equation}
The dominant intrinsic loss arises from Ohmic dissipation in the resonators. We model it by inserting a resistor $R$ into each resonator (Fig.~1a of the main text) and define the Rayleigh dissipation function in terms of the resonator currents: 
\begin{equation}
\begin{aligned}
    \mathcal{D}&=\sum_k\frac{R}{2}\dot{Q}_k^{L_r}\dot{Q}_{-k}^{L_r}=\sum_k\frac{R}{L'_{r,k}}\Phi_k^{L_r}\Phi_{-k}^{L_r}=-\sum_k \frac{R\omega_k}{4L'_{r,k}}(b_{-k}-b_k^\dagger)(b_k-b_{-k}^\dagger)\\
    &=\sum_k \mathbf{x}_k^\dagger\hat{D}_k\mathbf{x}_k,
\end{aligned}
\end{equation}
where the dissipation matrix is
\begin{equation}
    \hat{D}_k =\frac{R\omega_k}{4L'_{r,k}}\begin{bmatrix}
        0 & 0 & 0 & 0\\
        0 & 1 &  0 & -1\\
         0 &  0 & 0&0\\
         0& -1 & 0 &1\end{bmatrix}. 
\end{equation}
Then, the equation of motion for the mode vector  $\mathbf{x}_k$ is given by
\begin{equation}
\begin{aligned}
    i\frac{\partial\mathbf{x}_k}{\partial t}&=iM\hat{J}M^\dagger\frac{\partial\mathcal{H}}{\partial\mathbf{x}_k^\dagger}+iM\hat{R}M^\dagger\frac{\partial\mathcal{D}}{\partial\mathbf{x}_k^\dagger}\\
    &=iM\hat{J}M^\dagger2(\hat{H}_k^{(0)}+\hat{H}^{(1)}_k(t))\mathbf{x}_k+iM\hat{R}M^\dagger2\hat{D}_k\mathbf{x}_k\\
    &=\mathbf{H}_k(t)\mathbf{x}_k=(\mathbf{H}_k^{(0)}+\mathbf{H}^{(1)}_k(t)+\mathbf{D}_k)\mathbf{x}_k,
\end{aligned} \label{eqn:Hamilton Eqs}
\end{equation}
where
\begin{equation}
\begin{aligned}
&\mathbf{H}_k^{(0)} = \begin{bmatrix}
        \Omega_k & -g_k & 0 & -g_k\\
         -g_k & \omega_k &  -g_k & 0\\
         0 &  g_k & -\Omega_k&g_k\\
         g_k & 0 & g_k &-\omega_k\end{bmatrix}, \; \mathbf{H}^{(1)}_k(t)=\frac{\omega_k\delta(t)}{2}\begin{bmatrix}
        0 & 0 & 0 & 0\\
         0& 1 &  0 & 1\\
         0 &  0 &0&0\\
         0 & -1 & 0 &-1\end{bmatrix},    \; \mathrm{and}\\
         &\mathbf{D}_k = i\gamma\begin{bmatrix}
        0 & 0 & 0 & 0\\
         0& -1 &  0 & 1\\
         0 &  0 &0&0\\
         0 & 1 & 0 &-1\end{bmatrix}.
\end{aligned}
\end{equation}
Here, $\gamma\equiv R/(2L'_{r,k})$ denotes the intrinsic damping rate of the resonator Bloch mode. By Floquet's theorem, any solution to Eq.~\ref{eqn:Hamilton Eqs} can be expressed as $\mathbf{x}_k(t) = e^{-i\omega_\alpha t}\Psi_\alpha(t)$, where $\Psi_\alpha(t)$ is time-periodic with period $T$, i.e., $\Psi_\alpha(t+T)=\Psi_\alpha(t)$. Expanding the periodic part as a Fourier series, $\Psi_\alpha(t) = \sum_q e^{-iq\Omega t}\psi_k^q$, and substituting this into the equation of motion yields the Floquet eigenvalue equation
\begin{equation}
    \left[\mathbf{H}_k(t)-i\frac{\partial}{\partial t}\right]\Psi_\alpha(t) = \omega_\alpha\Psi_\alpha(t).\label{eqn:sm_eom_floquet}
\end{equation}
Because the operator in Eq.~\eqref{eqn:sm_eom_floquet} is time-periodic, if  $\omega_\alpha$ is an eigenfrequency, then so is  $\omega_\alpha+m\Omega$ for any integer $m$. This ladder of frequencies constitutes the quasi-frequency spectrum of the Floquet system. Substituting the Fourier series for $\Psi_\alpha(t)$ and $\mathbf{H}_k(t)=\sum_qe^{-iq\Omega t}\mathbf{H}_k^q$ into Eq.~\ref{eqn:sm_eom_floquet} yields the coupled relations
\begin{equation}
    (\omega_\alpha + q\Omega)\psi_{k,\alpha}^q=\sum_r\mathbf{H}_k^{q-r}\psi_{k,\alpha}^r,\
\end{equation}
which connect the Fourier coefficients $\psi_{k,\alpha}^q$ across different Floquet harmonics. This set of equations can be cast in matrix form as
\begin{equation}
    \mathbf{H}_k^F\mathbf{F}_{k,\alpha}=\omega_\alpha\mathbf{F}_{k,\alpha},
\end{equation}
where  $\mathbf{H}_k^F$ is the Floquet Hamiltonian,
\begin{equation}
    \mathbf{H}_k^F = \begin{bmatrix}
        \ddots & \ddots & & & \\
        \ddots & \mathbf{H}_k^0+\Omega I_4 & \mathbf{H}_k^{-1} & \mathbf{H}_k^{-2} & \\
               & \mathbf{H}_k^{+1} & \mathbf{H}_k^0 & \mathbf{H}_k^{-1} & \\
               &    \mathbf{H}_k^{+2}         & \mathbf{H}_k^{+1} & \mathbf{H}_k^0-\Omega I_4 & \ddots \\
               &    &   & \ddots & \ddots
    \end{bmatrix}, \label{eqn:Floquet_Ham} 
\end{equation}
and the corresponding Floquet eigenvector is $\mathbf{F}_{k,\alpha} = [\cdots;\psi_{k,\alpha}^{-1};\psi_{k,\alpha}^0;\psi_{k,\alpha}^1;\cdots]^T$. The off-diagonal blocks $\mathbf{H}_k^m \;(m\neq0)$ mediate the coupling between the original band, spanned by the eigenmodes of $\mathbf{H}_k^0$, and its frequency-shifted replicas, spanned by the eigenmodes of $\mathbf{H}_k^0+m\Omega I_4$. The eigenvalues $\omega_\alpha$ obtained from $\mathbf{H}_k^F$ therefore constitute the Floquet band structure of the photonic time crystal.

When the driving frequency is close to twice the resonator Bloch mode frequency, $\Omega\simeq2\omega_k$, only the positive‑frequency at $\omega_k$ band and the Floquet replica of its negative-frequency counterpart at $\Omega-\omega_k$ remain near resonance; all other sidebands are far detuned. In this scenario, the coupling between these two bands which gives rise to the formation of the momentum gap can be described by the following reduced $2\times2$ non-Hermitian Floquet Hamiltonian:
\begin{equation}
        \mathbf{H}_{k,\mathrm{red}}^F=\begin{bmatrix}
        -\omega_k+\Omega-i\gamma & -\frac{1}{4}\omega_k\delta\\
        \frac{1}{4}\omega_k\delta & \omega_k-i\gamma
        \end{bmatrix}.
\end{equation}
Here, we assume a single-harmonic modulation \(\delta(t) = \delta \cos(\Omega t)\). The eigenvalues of $\mathbf{H}_{k,\mathrm{red}}^F$ are given by
\begin{equation}
    \omega_\pm= \frac{\Omega}{2}-i\gamma \mp \sqrt{\bigg(\frac{\Omega}{2}-\omega_k\bigg)^2-\bigg(\frac{\omega_k\delta}{4}\bigg)^2}.
\end{equation}
Inside the momentum gap the square root is purely imaginary, so the two modes share the same real frequency $\Omega/2$ but acquire different damping rates. The critical modulation depth $\delta_c$ is defined as the point where the less-damped mode becomes lossless, i.e., when its imaginary part just reaches zero:
\begin{equation}
    \delta_c = \frac{4}{\omega_k}\sqrt{\Delta^2+\gamma^2}
\end{equation}
where $\Delta\equiv\frac{\Omega}{2}-\omega_k$ is the detuning from exact resonance.
For $\delta<\delta_c$ both modes remain lossy. When $\delta>\delta_c$ one mode acquires net positive gain, signaling the onset of a parametric instability, or PTC-laser (self-oscillation) transition.

\section{Power dissipation and local density of states in photonic time crystals}
We consider the case where the resonator at site $n_0$ is driven by an external voltage source $V_d(t)$. This adds the following term to the Lagrangian:
\begin{equation}
    \mathcal{L}_{drive} = Q_{n_0}^{L_r}V_d(t).
\end{equation}
All other terms in the Lagrangian remain as in Eq. \ref{eqn:Lag_position}. Here, the applied voltage \(V_d(t)\) drives a localized, oscillating dipole, i.e., the resonator charge $Q_{n_0}^{L_r}$ at site \(n_0\). Within linear response, the time-averaged power delivered by this dipole is proportional to the local density of states \cite{park2025spontaneous_SI}. Applying the inverse discrete Fourier transform to $Q_{n_0}^{L_r}$, the drive term becomes
\begin{equation}
\begin{aligned}
    \mathcal{L}_{drive} &=\frac{1}{\sqrt{N}}\sum_ke^{ikn_0}Q_k^{L_r}V_d(t)\\
    &=\frac{1}{\sqrt{N}}\sum_ke^{ikn_0}\sqrt{\frac{C_c\omega_k}{2}}(b_k+b_{-k}^\dagger)V_d(t).
\end{aligned}
\end{equation}
This term introduces an external force in the equation of motion for the mode vector $\mathbf{x}_k$:
\begin{equation}
    i\frac{\partial\mathbf{x}_k}{\partial t}=\mathbf{H}_k(t)\mathbf{x}_k+\mathbf{f}_k(t),
\end{equation}
where 
\begin{equation}
    \mathbf{f}_k(t)=\sqrt{\frac{C_c\omega_k}{2N}}e^{ikn_0}V_d(t)\begin{bmatrix}
        0 \\
         -1 \\
         0 \\
         1 \end{bmatrix}.
\end{equation}
We take the drive to be monochromatic, 
\begin{equation}
    V_d(t) = V_0e^{-i\omega t}+V_0^*e^{i\omega t}.
\end{equation}
Retaining only the positive-frequency component $V_0e^{-i\omega t}$, the driven Floquet equation reads
\begin{equation}
\omega\mathbf{F}_k=\mathbf{H}_k^F\mathbf{F}_k + V_0f_k\mathbf{s},
\end{equation}
with
\begin{equation}
    f_k = \sqrt{\frac{C_c \omega_k}{2N}}\, e^{i k n_0}, \quad
\mathbf{s} = [\cdots;\, \mathbf{0};\, [0, -1, 0, 1]^T;\, \mathbf{0};\, \cdots],
\end{equation}
where $ \mathbf{s} $ has non-zero entries only for Floquet index $q = 0$. Solving for $\mathbf{F}_k$ gives
\begin{equation}
    \mathbf{F}_k = (\omega \mathbf{I}_F-\mathbf{H}_k^F)^{-1}f_k\mathbf{s}\equiv G_F(k,\omega)f_k\mathbf{s}.
\end{equation}
with $G_F(k,\omega)$ the Floquet Green's function. The resonator-current component oscillating at frequency $\omega$ (i.e., with time dependence $e^{-i\omega t }$, corresponding to the  $q=0$ Floquet index) is given by 
\begin{equation}
    \tilde{\dot{Q}}_{k}^{L_r,0} =\frac{iV_0}{2\sqrt{N}L'_{r,k}}e^{-ikn_0}\mathbf{s}^TG_F(k,\omega)\mathbf{s}\equiv\frac{iV_0}{2\sqrt{N}L'_{r,k}}e^{-ikn_0}g_F(k,\omega),
\end{equation}
where we define $ g_F(k,\omega)\equiv\mathbf{s}^TG_F(k,\omega)\mathbf{s}$. The conjugate drive component $V_0^*e^{i\omega t}$ produces the current
\begin{equation}
    \tilde{\tilde{\dot{Q}}}_{k}^{L_r,0} =\frac{iV_0^*}{2\sqrt{N}L'_{r,k}}e^{-ikn_0}g_F(k,-\omega).
\end{equation}
Hence, the time-averaged power delivered by the source is
\begin{equation}
\begin{aligned}
    \bar{P} &= -\left\langle V_d(t) \cdot \dot{Q}_{n_0}^{L_r}(t) \right\rangle \\
    &= -\left\langle \left( V_0 e^{-i \omega t} + V_0^* e^{i \omega t} \right)
    \left( \frac{1}{\sqrt{N}} \sum_k e^{i k n_0} \dot{Q}_k^{L_r}(t) \right) \right\rangle \\
    &= -\frac{1}{\sqrt{N}} \sum_k e^{i k n_0} \left( V_0^* \tilde{\dot{Q}}_k^{L_r,0} + V_0 \tilde{\tilde{\dot{Q}}}_{k}^{L_r,0} \right) \\
    &= -\frac{|V_0|^2}{N} \sum_k \frac{i}{2 L'_{r,k}} \left[ g_F(k, \omega) + g_F(k, -\omega) \right] \\
    &= \frac{|V_0|^2}{N} \sum_k \frac{1}{L'_{r,k}}\, \Im \left[ g_F(k,\omega) \right].
\end{aligned} \label{eqn:Power} 
\end{equation}
Since $g_F(k,\omega) $ is the diagonal element of the Floquet Green’s function projected onto the driven resonator, the time-averaged power delivered by the source (and hence absorbed by the circuit in steady state) is directly proportional to the LDOS at site $n_0$. Specifically, we define the LDOS as 
\begin{equation}
\rho(\omega;n_{0})
=\frac{2}{\pi}\,\frac{1}{N}\sum_{k}
     \frac{1}{L'_{r,k}}\,
     \Im\bigl[g_{F}(k,\omega)\bigr] . \label{eqn:LDOS} 
\end{equation}
To resolve the LDOS contribution at each momentum $k$, we define the momentum-resolved density of states (kDOS) as 
\begin{equation}
    \rho_k(\omega) \equiv \frac{2}{\pi}\frac{1}{L'_{r,k}}\,
     \Im\bigl[g_{F}(k,\omega)\bigr]
\end{equation}
such that the total LDOS satisfies $\rho(\omega;n_0)=N^{-1}\sum_k\rho_k(\omega)$.
Combining Eqs. \ref{eqn:Power} and \ref{eqn:LDOS} , we obtain a compact expression that relates the dissipated power to the LDOS: 
\begin{equation}
 \bar{P}=\frac{\pi}{2}\,|V_{0}|^{2}\,
        \rho(\omega;n_{0}) .
\end{equation}
This result highlights that the LDOS governs how efficiently external energy is absorbed by the system at frequency $\omega$. In experimental measurements, the observed spectrum may include additional contributions from system noise sources such as thermal fluctuations or amplifier noise. Nonetheless, the measured spectral response still reflects features of the underlying LDOS, allowing its profile to be inferred from experimental data.

\section{Experimental Implementation of the PTCs}
\begin{figure*}
  \centering
    \includegraphics[width=0.9\textwidth]{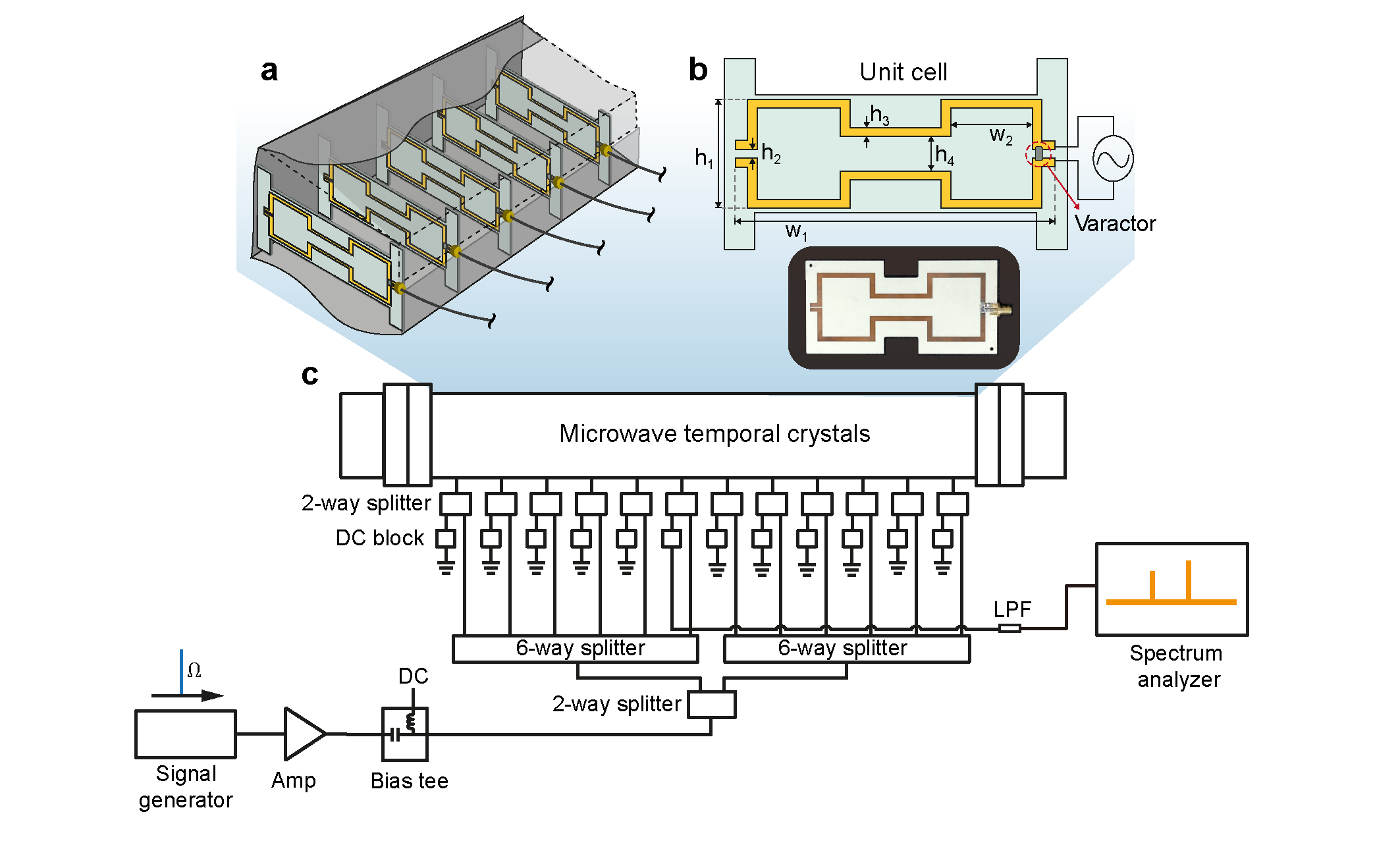}
   \caption[Experimental realization of the photonic time crystal]{\label{fig:SI_fig1}(a) Schematic of the one-dimensional array of 12 coupled LC resonators embedded in a waveguide. (b) Schematic layout and photograph of the fabricated unit cell, incorporating a varactor diode for time-periodic capacitance modulation via a DC-biased AC voltage. The unit cell is designed with the following geometrical parameters: $\mathrm{h}_1=40\ \mathrm{mm}$, $\mathrm{h}_2=1\ \mathrm{mm}$, $\mathrm{h}_3=3\ \mathrm{mm}$, $\mathrm{h}_4=10\ \mathrm{mm}$, $\mathrm{w}_1=105\ \mathrm{mm}$ and $\mathrm{w}_2=27.5\ \mathrm{mm}$. (c) Schematic of the experimental setup.
 }
\end{figure*}

To implement the unit cell of the LC resonator, a metallic split-ring resonator (SRR) was patterned onto a printed circuit board (Fig.~\ref{fig:SI_fig1}b). Each unit cell incorporates a varactor diode (SMV1247, Skyworks), whose capacitance can be modulated by applying a DC-biased AC voltage across it. As a result, the resonant characteristics of the unit cell undergo time-periodic modulation. In the experiment, a DC bias voltage of 2.6\,V was applied. To construct a one-dimensional array of coupled LC resonators, 12 unit cells were placed inside a waveguide (Fig.~\ref{fig:SI_fig1}a). 
The unit cells were arranged with a center-to-center spacing of 10\,mm and the waveguide was terminated at both ends with 50\;$\Omega$ loads to suppress reflections.
The complete experimental setup is shown in Fig.~\ref{fig:SI_fig1}c.

The modulation signal (AC voltage) is generated by a signal generator (N5171B, Keysight) and amplified by a
radio-frequency amplifier (RUM43020-10, RFHIC). The amplified AC voltage is then DC-biased using a bias tee and subsequently distributed through a two-way splitter, followed by a six-way splitter. Additional two-way splitters are inserted just before the resonator array, with each splitter enabling individual signal measurement from a corresponding unit cell. To prevent unwanted DC current flow, a DC block is inserted at one end of the two-way splitter. Additionally, a low-pass filter (LPF) is placed before the oscilloscope (DPO 70804, Tektronix) or spectrum analyzer (E4404B, Agilent) to suppress reflections of the modulation signal.

To examine the predicted enhancement of the LDOS near the momentum-gap frequency, we operate the structure with no coherent input; the only drive is broadband white noise arising from Johnson–Nyquist thermal fluctuations in combination with residual technical noise. These fluctuations act as spatially distributed, broadband noise sources that probe the linear response of the PTC. The measured noise-driven radiated spectrum closely follows the LDOS lineshape derived from the non-Hermitian Floquet framework, providing an experimental proxy for the frequency dependence of the spontaneous-emission rate in the weak-coupling limit.
\putbib[refs_SI]
\clearpage
\end{widetext}
\end{bibunit}


\providecommand{\noopsort}[1]{}\providecommand{\singleletter}[1]{#1}%
\begin{thebibliography}{10}

\bibitem{PhysRev.69.674.2}
Edward~Mills Purcell.
\newblock Spontaneous emission probabilities at radio frequencies.
\newblock {\em Phys. Rev.}, 69:674--674, Jun 1946.

\bibitem{1972JETP...35..269B}
V.~P. {Bykov}.
\newblock {Spontaneous Emission in a Periodic Structure}.
\newblock {\em Soviet Journal of Experimental and Theoretical Physics}, 35:269, January 1972.

\bibitem{PhysRevLett.58.2059}
Eli Yablonovitch.
\newblock Inhibited spontaneous emission in solid-state physics and electronics.
\newblock {\em Phys. Rev. Lett.}, 58:2059--2062, May 1987.

\bibitem{john1987strong}
Sajeev John.
\newblock Strong localization of photons in certain disordered dielectric superlattices.
\newblock {\em Physical review letters}, 58(23):2486, 1987.

\bibitem{fan1997high}
Shanhui Fan, Pierre~R Villeneuve, JD~Joannopoulos, and EF~Schubert.
\newblock High extraction efficiency of spontaneous emission from slabs of photonic crystals.
\newblock {\em Physical review letters}, 78(17):3294, 1997.

\bibitem{PhysRevLett.95.013904}
Dirk Englund, David Fattal, Edo Waks, Glenn Solomon, Bingyang Zhang, Toshihiro Nakaoka, Yasuhiko Arakawa, Yoshihisa Yamamoto, and Jelena Vu\ifmmode \check{c}\else \v{c}\fi{}kovi\ifmmode~\acute{c}\else \'{c}\fi{}.
\newblock Controlling the spontaneous emission rate of single quantum dots in a two-dimensional photonic crystal.
\newblock {\em Phys. Rev. Lett.}, 95:013904, Jul 2005.

\bibitem{PhysRevLett.99.023902}
G.~Lecamp, P.~Lalanne, and J.~P. Hugonin.
\newblock Very large spontaneous-emission $\ensuremath{\beta}$ factors in photonic-crystal waveguides.
\newblock {\em Phys. Rev. Lett.}, 99:023902, Jul 2007.

\bibitem{PhysRevLett.99.193901}
V.~S. C.~Manga Rao and S.~Hughes.
\newblock Single quantum dot spontaneous emission in a finite-size photonic crystal waveguide: Proposal for an efficient ``on chip'' single photon gun.
\newblock {\em Phys. Rev. Lett.}, 99:193901, Nov 2007.

\bibitem{Noda2007}
Susumu Noda, Masayuki Fujita, and Takashi Asano.
\newblock Spontaneous-emission control by photonic crystals and nanocavities.
\newblock {\em Nature Photonics}, 1(8):449--458, Aug 2007.

\bibitem{jacob2012broadband}
Zubin Jacob, Igor~I Smolyaninov, and Evgenii~E Narimanov.
\newblock Broadband purcell effect: Radiative decay engineering with metamaterials.
\newblock {\em Applied Physics Letters}, 100(18), 2012.

\bibitem{noginov2010controlling}
MA~Noginov, H~Li, Yu~A Barnakov, D~Dryden, G~Nataraj, G~Zhu, CE~Bonner, M~Mayy, Z~Jacob, and EE~Narimanov.
\newblock Controlling spontaneous emission with metamaterials.
\newblock {\em Optics letters}, 35(11):1863--1865, 2010.

\bibitem{lodahl2004controlling}
Peter Lodahl, A~Floris~van Driel, Ivan~S Nikolaev, Arie Irman, Karin Overgaag, Dani{\"e}l Vanmaekelbergh, and Willem~L Vos.
\newblock Controlling the dynamics of spontaneous emission from quantum dots by photonic crystals.
\newblock {\em Nature}, 430(7000):654--657, 2004.

\bibitem{vos2009orientation}
Willem~L Vos, A~Femius Koenderink, and Ivan~S Nikolaev.
\newblock Orientation-dependent spontaneous emission rates of a two-level quantum emitter in any nanophotonic environment.
\newblock {\em Physical Review A—Atomic, Molecular, and Optical Physics}, 80(5):053802, 2009.

\bibitem{novotny2012principles}
Lukas Novotny and Bert Hecht.
\newblock {\em Principles of nano-optics}.
\newblock Cambridge university press, 2012.

\bibitem{milonni2013quantum}
Peter~W Milonni.
\newblock {\em The quantum vacuum: an introduction to quantum electrodynamics}.
\newblock Academic press, 2013.

\bibitem{zurita2009reflection}
Jorge~R Zurita-S{\'a}nchez, P~Halevi, and Juan~C Cervantes-Gonzalez.
\newblock Reflection and transmission of a wave incident on a slab with a time-periodic dielectric function.
\newblock {\em Physical Review A—Atomic, Molecular, and Optical Physics}, 79(5):053821, 2009.

\bibitem{salem2015temporal}
MA~Salem and Christophe Caloz.
\newblock Temporal photonic crystals: Causality versus periodicity.
\newblock In {\em 2015 International Conference on Electromagnetics in Advanced Applications (ICEAA)}, pages 490--493. IEEE, 2015.

\bibitem{PhysRevA.93.063813}
Juan~Sabino Mart\'{\i}nez-Romero, O.~M. Becerra-Fuentes, and P.~Halevi.
\newblock Temporal photonic crystals with modulations of both permittivity and permeability.
\newblock {\em Phys. Rev. A}, 93:063813, Jun 2016.

\bibitem{doi:10.1063/1.4928659}
J.~R. Reyes-Ayona and P.~Halevi.
\newblock Observation of genuine wave vector (k or beta) gap in a dynamic transmission line and temporal photonic crystals.
\newblock {\em Applied Physics Letters}, 107(7):074101, 2015.

\bibitem{PhysRevB.98.085142}
Neng Wang, Zhao-Qing Zhang, and C.~T. Chan.
\newblock Photonic floquet media with a complex time-periodic permittivity.
\newblock {\em Phys. Rev. B}, 98:085142, Aug 2018.

\bibitem{martinez2018parametric}
Juan~Sabino Mart{\'\i}nez-Romero and P~Halevi.
\newblock Parametric resonances in a temporal photonic crystal slab.
\newblock {\em Physical Review A}, 98(5):053852, 2018.

\bibitem{Chamanara2018}
Nima Chamanara, Zo{\'{e}}~Lise Deck-L{\'{e}}ger, Christophe Caloz, and Dikshitulu Kalluri.
\newblock {Unusual electromagnetic modes in space-time-modulated dispersion-engineered media}.
\newblock {\em Physical Review A}, 97(6), jun 2018.

\bibitem{Park2021}
Jagang Park and Bumki Min.
\newblock {Spatiotemporal plane wave expansion method for arbitrary space–time periodic photonic media}.
\newblock {\em Optics Letters}, 46(3):484--487, 2021.

\bibitem{Lee2021}
Seojoo Lee, Jagang Park, Hyukjoon Cho, Yifan Wang, Brian Kim, Chiara Daraio, and Bumki Min.
\newblock {Parametric oscillation of electromagnetic waves in momentum band gaps of a spatiotemporal crystal}.
\newblock {\em Photonics Research}, 9(2):142, 2021.

\bibitem{asgari2024theory_}
Mohammad~M Asgari, Puneet Garg, Xuchen Wang, Mohammad~S Mirmoosa, Carsten Rockstuhl, and Viktar Asadchy.
\newblock Theory and applications of photonic time crystals: a tutorial.
\newblock {\em Advances in Optics and Photonics}, 16(4):958--1063, 2024.

\bibitem{wang2024expanding}
X.~Wang, P.~Garg, M.~S. Mirmoosa, A.~G. Lamprianidis, C.~Rockstuhl, and V.~S. Asadchy.
\newblock Expanding momentum bandgaps in photonic time crystals through resonances.
\newblock {\em Nature Photonics}, Nov 2024.

\bibitem{sustaeta2025quantum}
Jaime~E Sustaeta-Osuna, Francisco~J Garcia-Vidal, and PA~Huidobro.
\newblock Quantum theory of photon pair creation in photonic time crystals.
\newblock {\em ACS Photonics}, 2025.

\bibitem{dong2025extremely}
Zhaohui Dong, Xianfeng Chen, and Luqi Yuan.
\newblock Extremely narrow band in moir{\'e} photonic time crystal.
\newblock {\em Physical Review Letters}, 135(3):033803, 2025.

\bibitem{doi:10.1126/sciadv.abo6220}
Jagang Park, Hyukjoon Cho, Seojoo Lee, Kyungmin Lee, Kanghee Lee, Hee~Chul Park, Jung-Wan Ryu, Namkyoo Park, Sanggeun Jeon, and Bumki Min.
\newblock Revealing non-hermitian band structure of photonic floquet media.
\newblock {\em Science Advances}, 8(40):eabo6220, 2022.

\bibitem{doi:10.1126/sciadv.adg7541}
Xuchen Wang, Mohammad~Sajjad Mirmoosa, Viktar~S. Asadchy, Carsten Rockstuhl, Shanhui Fan, and Sergei~A. Tretyakov.
\newblock Metasurface-based realization of photonic time crystals.
\newblock {\em Science Advances}, 9(14):eadg7541, 2023.

\bibitem{feinberg2025plasmonic}
Joshua Feinberg, David~E Fernandes, Boris Shapiro, and M{\'a}rio~G Silveirinha.
\newblock Plasmonic time crystals.
\newblock {\em Physical Review Letters}, 134(18):183801, 2025.

\bibitem{park2025spontaneous}
Jagang Park, Kyungmin Lee, Ruo-Yang Zhang, Hee-Chul Park, Jung-Wan Ryu, Gil~Young Cho, Min~Yeul Lee, Zhaoqing Zhang, Namkyoo Park, Wonju Jeon, et~al.
\newblock Spontaneous emission decay and excitation in photonic time crystals.
\newblock {\em Physical Review Letters}, 135(13):133801, 2025.

\bibitem{bae2025cavity}
Junhyeon Bae, Kyungmin Lee, Bumki Min, and Kun~Woo Kim.
\newblock Cavity quantum electrodynamics of photonic temporal crystals.
\newblock {\em arXiv preprint arXiv:2501.03106}, 2025.

\bibitem{eckardt1991optical}
Robert~C Eckardt, CD~Nabors, William~J Kozlovsky, and Robert~L Byer.
\newblock Optical parametric oscillator frequency tuning and control.
\newblock {\em Journal of the Optical Society of America B}, 8(3):646--667, 1991.

\bibitem{yao2020classical}
Norman~Y Yao, Chetan Nayak, Leon Balents, and Michael~P Zaletel.
\newblock Classical discrete time crystals.
\newblock {\em Nature Physics}, 16(4):438--447, 2020.

\bibitem{PhysRevLett.123.124301}
Toni~L. Heugel, Matthias Oscity, Alexander Eichler, Oded Zilberberg, and R.~Chitra.
\newblock Classical many-body time crystals.
\newblock {\em Phys. Rev. Lett.}, 123:124301, Sep 2019.

\bibitem{pizzi2021aclassical}
Andrea Pizzi, Andreas Nunnenkamp, and Johannes Knolle.
\newblock Classical approaches to prethermal discrete time crystals in one, two, and three dimensions.
\newblock {\em Physical Review B}, 104(9):094308, 2021.

\bibitem{PhysRevE.100.060105}
F.~M. Gambetta, F.~Carollo, A.~Lazarides, I.~Lesanovsky, and J.~P. Garrahan.
\newblock Classical stochastic discrete time crystals.
\newblock {\em Phys. Rev. E}, 100:060105, Dec 2019.

\bibitem{taheri2022all}
Hossein Taheri, Andrey~B Matsko, Lute Maleki, and Krzysztof Sacha.
\newblock All-optical dissipative discrete time crystals.
\newblock {\em Nature communications}, 13(1):848, 2022.

\bibitem{giergiel2023discrete}
Krzysztof Giergiel, Jia Wang, Bryan~J Dalton, Peter Hannaford, and Krzysztof Sacha.
\newblock Discrete time crystals with absolute stability.
\newblock {\em Physical Review B}, 108(18):L180201, 2023.

\bibitem{PhysRevLett.127.140603}
Bingtian Ye, Francisco Machado, and Norman~Y. Yao.
\newblock Floquet phases of matter via classical prethermalization.
\newblock {\em Phys. Rev. Lett.}, 127:140603, Sep 2021.

\bibitem{yi2024theory}
Stuart Yi-Thomas and Jay~D Sau.
\newblock Theory for dissipative time crystals in coupled parametric oscillators.
\newblock {\em Physical Review Letters}, 133(26):266601, 2024.

\bibitem{choi2017observation}
Soonwon Choi, Joonhee Choi, Renate Landig, Georg Kucsko, Hengyun Zhou, Junichi Isoya, Fedor Jelezko, Shinobu Onoda, Hitoshi Sumiya, Vedika Khemani, et~al.
\newblock Observation of discrete time-crystalline order in a disordered dipolar many-body system.
\newblock {\em Nature}, 543(7644):221--225, 2017.

\bibitem{zhang2017observation}
Jiehang Zhang, Paul~W Hess, A~Kyprianidis, Petra Becker, A~Lee, J~Smith, Gaetano Pagano, I-D Potirniche, Andrew~C Potter, Ashvin Vishwanath, et~al.
\newblock Observation of a discrete time crystal.
\newblock {\em Nature}, 543(7644):217--220, 2017.

\bibitem{ho2017critical}
Wen~Wei Ho, Soonwon Choi, Mikhail~D Lukin, and Dmitry~A Abanin.
\newblock Critical time crystals in dipolar systems.
\newblock {\em Physical review letters}, 119(1):010602, 2017.

\bibitem{rovny2018p}
Jared Rovny, Robert~L Blum, and Sean~E Barrett.
\newblock P 31 nmr study of discrete time-crystalline signatures in an ordered crystal of ammonium dihydrogen phosphate.
\newblock {\em Physical Review B}, 97(18):184301, 2018.

\bibitem{liu2023photonic}
Tongjun Liu, Jun-Yu Ou, Kevin~F MacDonald, and Nikolay~I Zheludev.
\newblock Photonic metamaterial analogue of a continuous time crystal.
\newblock {\em Nature Physics}, 19(7):986--991, 2023.

\bibitem{kyprianidis2021observation}
Antonis Kyprianidis, Francisco Machado, William Morong, Patrick Becker, Kate~S Collins, Dominic~V Else, Lei Feng, Paul~W Hess, Chetan Nayak, Guido Pagano, et~al.
\newblock Observation of a prethermal discrete time crystal.
\newblock {\em Science}, 372(6547):1192--1196, 2021.

\bibitem{mi2022time}
Xiao Mi, Matteo Ippoliti, Chris Quintana, Ami Greene, Zijun Chen, Jonathan Gross, Frank Arute, Kunal Arya, Juan Atalaya, Ryan Babbush, et~al.
\newblock Time-crystalline eigenstate order on a quantum processor.
\newblock {\em Nature}, 601(7894):531--536, 2022.

\end{thebibliography}


\begin{thebibliography}{1}

\bibitem{park2022revealing}
Jagang Park, Hyukjoon Cho, Seojoo Lee, Kyungmin Lee, Kanghee Lee, Hee~Chul Park, Jung-Wan Ryu, Namkyoo Park, Sanggeun Jeon, and Bumki Min.
\newblock Revealing non-hermitian band structure of photonic floquet media.
\newblock {\em Science advances}, 8(40):eabo6220, 2022.

\bibitem{park2025spontaneous_SI}
Jagang Park, Kyungmin Lee, Ruo-Yang Zhang, Hee-Chul Park, Jung-Wan Ryu, Gil~Young Cho, Min~Yeul Lee, Zhaoqing Zhang, Namkyoo Park, Wonju Jeon, et~al.
\newblock Spontaneous emission decay and excitation in photonic time crystals.
\newblock {\em Physical Review Letters}, 135(13):133801, 2025.

\end{thebibliography}
\end{document}